\documentclass[aps,prd,preprint,superscriptaddress]{revtex4}

\usepackage{bm}
\usepackage{amssymb,graphicx,amsmath,array,verbatim,amscd}

\def\ee{\end{eqnarray}}

\def\p{\partial}

\def\D{\mathcal{D}}

\def\=:{=\hspace{-.7em}\raisebox{1.1ex}{.}\hspace{.1em}\raisebox{-0.2ex}{.} }

\newcommand {\beq}{\begin{eqnarray}}
\newcommand {\eeq}{\end{eqnarray}}

\newcommand {\tr}{{\rm tr}\,}

\newcommand{\hs}[1]{\hspace{#1 mm}}

\newcommand{\R}{\mathbb{R}}
\newcommand{\C}{\mathbb{C}}
\newcommand{\ba}{\left( \begin{array}}
\newcommand{\ea}{\end{array} \right)}
\newcommand{\bea}{\begin{eqnarray}}
\newcommand{\eea}{\end{eqnarray}}
\newcommand{\beann}{\begin{eqnarray*}}
\newcommand{\eeann}{\end{eqnarray*}}
\begin{document}


\title{All Exact Solutions of Non-Abelian 
Vortices\\ 
from Yang-Mills Instantons
}


\author{Minoru Eto}
\affiliation{
Department of Physics, Yamagata University, Yamagata 990-8560, Japan
}

\author{Toshiaki Fujimori}
\affiliation{
Department of Physics, National Taiwan University, Taipei 10617, Taiwan
}

\author{Muneto Nitta}
\affiliation{
Department of Physics, and Research and Education Center for Natural 
Sciences, Keio University, Hiyoshi 4-1-1, Yokohama, Kanagawa 223-8521, Japan
}

\author{Keisuke Ohashi}
\affiliation{
Department of Mathematics and Physics, Graduate School of Science, Osaka City University,
3-3-138 Sugimoto, Sumiyoshi, Osaka, 558-8585, Japan
}



\date{\today}
\begin{abstract}
We successfully exhaust the complete set of exact solutions 
of non-Abelian vortices in a quiver gauge theory, that is, 
the $S[U(N)\times U(N)]$ gauge theory with a bi-fudamental scalar field on a hyperbolic plane with a certain curvature, 
from $SO(3)$-invariant $SU(2N)$ Yang-Mills instanton solutions. 
This work provides, for the first time, exact non-Abelian vortex solutions. 
We establish the ADHM construction for non-Abelian vortices 
and identify all the moduli parameters and the complete moduli space.

\end{abstract}
\pacs{14.80.Hv, 11.27.+d, 12.10.-g, 11.30.Pb}

\maketitle

\section{Introduction} 

Since the discovery of non-Abelian vortices 
\cite{Hanany:2003hp,Auzzi:2003fs,Eto:2005yh}, 
they have been studied extensively \cite{Tong:2005un,Eto:2006pg,Shifman:2007ce}. While they are a natural extension of Abelian vortices 
\cite{Abrikosov:1956sx,Nielsen:1973cs} 
appearing in conventional superconductors, 
their analogues also appear in high-density quantum chromodynamics 
(QCD) showing color superconductivity  \cite{Balachandran:2005ev}.
In supersymmetric gauge theories, they are Bogomol'nyi-Prasad-Sommerfield (BPS) solitons \cite{Bogomolny:1975de} and are stable not only classically but also perturbatively and non-perturbatively. 
BPS non-Abelian vortices serve as an elegant tool to demonstrate \cite{Shifman:2004dr} 
the coincidence of BPS spectra in four-dimensional gauge theories 
and two-dimensional sigma models \cite{Dorey:1998yh}. 
Non-Abelian vortices also play prominent roles 
as instantons in non-perturbative dynamics of gauge theories in lower dimensions, 
similar to the role of Yang-Mills instantons \cite{Belavin:1975fg} 
in four dimensions; 
the non-perturbative partition function has been extensively studied 
by the vortex counting in ${\cal N}=(2,2)$ supersymmetric gauge theories in 
two dimensions \cite{Shadchin:2006yz}, 
similar to the instanton counting in four dimensions \cite{Nekrasov:2002qd}. 

However, vortex equations are not integrable even in the BPS limit 
\cite{Inami:2006wr}, 
and explicit solutions and the moduli space metric are not available. 
This is in contrast to the case of the self-dual equations 
for Yang-Mills instantons, 
for which the well-known Atiyah-Drinfeld-Hitchin-Manin (ADHM) construction  
 is available \cite{Atiyah:1978ri}. 
Thus far, some efforts to obtain the moduli space have been made.
The moduli space of non-Abelian vortices 
was determined without the moduli space metric 
by solving half of the BPS equations \cite{Eto:2005yh}.
The moduli space metric  
was obtained implicitly with a matrix function satisfying 
a differential equation \cite{Eto:2006uw}. 
The asymptotic metric for well-separated vortices was 
obtained \cite{Fujimori:2010fk} to study the low-energy scattering \cite{Eto:2011pj}.
The metrics were also obtained on submanifolds 
for the coincidence limit \cite{Eto:2006db} 
and on the symmetry orbits \cite{Eto:2010aj}.
However, the full moduli space is far beyond our reach
because of the non-integrability. 

Nevertheless, with changes to the geometry, 
the situation can become totally different. 
The BPS Abelian vortex equations on the hyperbolic plane ${\mathbb H}^2$ 
of curvature $-1/2$ are integrable \cite{Witten:1976ck}, 
and a general formula for the exact moduli space metric has been obtained
\cite{Strachan:1992fb}. 
The integrability is a consequence of the fact that 
these vortices are obtained as a dimensional reduction 
from $SO(3)$-symmetric Yang-Mills instantons on flat space ${\mathbb R}^4$ \cite{Witten:1976ck}.
More generally, the vortex equations 
on Riemann surfaces $\Sigma$ are integrable, 
when they are obtained from self-dual Yang-Mills equations on 
$\Sigma \times S^2$ \cite{Popov:2007ms}. 
Recently, hyperbolic vortices have been 
studied extensively \cite{Krusch:2009tn}. 
In particular, 
BPS non-Abelian vortex equations in a quiver gauge theory, {\it i.e.}, 
$S[U(N) \times U(N)]$ gauge theory coupled with 
a bi-fundamental scalar field on a hyperbolic space 
were obtained from $SO(3)$-symmetric 
$SU(2N)$ Yang-Mills instantons in a previous study \cite{Manton:2010wu}. 
However,  that study only considered 
 embedding of the Abelian vortex solutions into the diagonal $U(1)^N$ subgroup.
The same vortices on a flat space 
were also studied in another work \cite{Popov:2005ik}.
However, there remain open question on whether 
these vortices have non-trivial orientational moduli 
or what the complete set of solutions is.  

In this Letter,  
we construct, for the first time, a complete set of 
all the exact solutions of non-Abelian vortices 
in the $S[U(N)\times U(N)]$ gauge theory with a bi-fundamental scalar field 
on a hyperbolic plane with a certain curvature. 
We use $SO(3)$-invariant $SU(2N)$ Yang-Mills instanton solutions. 
We also establish the ADHM construction for non-Abelian vortices 
and identify all the moduli parameters and the complete moduli space. 


\section{Hyperbolic Vortices from $SO(3)$-Invariant Instantons}
\label{sec:vortex-instanton}

\subsection{$S[U(N) \times U(N)]$ vortices on a hyperbolic plane}

We consider a hyperbolic plane ${\mathbb H}^2$ as the upper half plane with a complex coordinate $z$, with $r \equiv {\rm Im} \, z>0$, endowed with the metric 
\beq
 g_{z \bar z} ~=~ - \frac{2 R^2}{(z-\bar z)^2} ~=~ \frac{R^2}{2 r^2}.
\eeq
The constant $R$ is related to 
the scalar curvature $-1/R^2$. See Appendix \ref{sec:hyperbolic}.

Let us consider the $U(N) \times U(N)$ gauge theory 
with gauge fields $A_z(z,\bar z)$ and $\tilde A_z(z,\bar z)$, 
coupled with a single bi-fundamental Higgs field $H(z,\bar z)$. 
Note that the overall $U(1)$ gauge group is trivial, 
and hence, the actual gauge group is $S[U(N) \times U(N)]$. 
For simplicity, we take the gauge coupling $g$ to be 
common for all gauge groups. 
The covariant derivative is $\D_{\bar z} H = (\p_{\bar z} + i A_{\bar z} H - i H \tilde A_{\bar z})$. In this setup, the action of our model is expressed as 
\beq
S = v^2 \int d^2 x \, \sigma \, \tr \left[ \frac{1}{\sigma^2} |F_{z \bar z}|^2 
+ \frac{1}{\sigma^2} |\tilde F_{z \bar z}|^2 
+ \frac{2}{\sigma} |\D_z H|^2 + \frac{2}{\sigma} |\D_{\bar z} H|^2 
\right. \nonumber\\
\left.
+ {\lambda \over 4} (HH^\dagger - \mathbf 1_N)^2 
+ {\lambda \over 4} (H^\dagger H - \mathbf 1_N)^2 \right],
\label{eq:action}
\eeq
where the Higgs field $H$ is rescaled so that its vacuum expectation value $v$ becomes the overall constant of the Lagrangian. The function $\sigma$ is the rescaled hyperbolic metric defined by
\beq
\sigma ~\equiv~ \frac{g^2 v^2}{2} g_{z \bar z} ~=~ \frac{g^2 v^2 R^2}{4r^2}. \label{eq:sigma}
\eeq
In this Letter, we consider the critical coupling 
(the BPS limit)
${\lambda \over 4} = 1$ 
and the ``integrable" case:
\beq
R = \frac{1}{g v} ~~~\Longleftrightarrow~~~ \sigma = \frac{1}{4r^2}.
\label{eq:integrable}
\eeq
The action in Eq.~(\ref{eq:action}) can be 
rewritten in the following form:
\beq
E &=& v^2 \int d^2 x \, 
\tr \bigg[ \sigma \left| i \sigma^{-1} F_{z \bar z} 
+ HH^\dagger - \mathbf 1_N \right|^2 
+ \sigma \left| i \sigma^{-1} \tilde F_{z \bar z} 
- H^\dagger H + \mathbf 1_N \right|^2 + 4 |\D_{\bar z} H|^2 \notag \\
&{}& \hs{25} + 2 \D_{\bar z} (\D_z H H^\dagger) 
- 2 \D_z (\D_{\bar z} H H^\dagger) 
+ 2 i (F_{z \bar z} - \tilde F_{z \bar z}) \bigg].
\eeq
Since the covariant derivative of the scalar field $\D_\mu H$ should vanish at the boundary of the hyperbolic plane, the lower bound of the action is given by
\beq
S \geq - v^2 \int_{\mathbb H} \tr (F - \tilde F).
\eeq
This Bogomol'nyi bound is saturated if 
the following BPS vortex equations are satisfied
\beq
0 &=& \D_{\bar z} H, 
\label{eq:BPSeq1}\\
0 &=& i \sigma^{-1} F_{z \bar z} + HH^\dagger - \mathbf 1_N, 
\label{eq:BPSeq2}\\ 
0 &=& i \sigma^{-1} \tilde F_{z \bar z} - H^\dagger H + \mathbf 1_N.
\label{eq:BPSeq3}
\eeq

\subsection{$SU(2N)$ instanton to $S[U(N) \times U(N)]$ vortices} 
\begin{figure}[h]
\begin{center}
\includegraphics[width=0.3\linewidth,keepaspectratio]{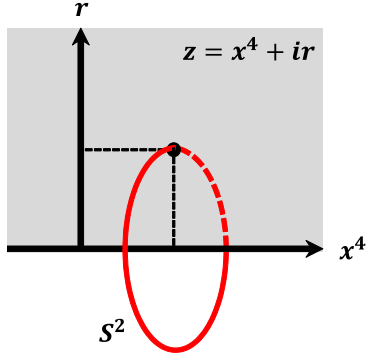}
\end{center}
\caption{\label{fig:SO(3)}
We consider the $SO(3)$ action on $(x^1,x^2,x^3)$.
The $x^4$ axis is a fixed line.
The grey region is an upper-half plane or a hyperbolic surface 
${\mathbb H}^2$. 
} 
\end{figure}

Here, we consider the $SO(3)$-rotationally-invariant 
$SU(2N)$ Yang-Mills instantons in
four-dimensional Euclidean space ${\mathbb R}^4$. 
The $SO(3)$ action on ${\mathbb R}^4$ 
rotates the coordinates $(x_1,x_2,x_3)$ and 
leaves the $x^4$-axis as fixed, 
while the $SO(3)$ orbit of a point is $S^2$, 
as shown in Fig.~\ref{fig:SO(3)}.
We obtain an upper-half plane ${\mathbb H}^2$
by an $S^2$-dimensional reduction from ${\mathbb R}^4$ 
with the $SO(3)$ fixed line (the $x^4$-axis) removed.
Since the classical pure Yang-Mills theory is conformally invariant, 
the conformal equivalence, 
$ {\mathbb R}^4 - {\mathbb R} \sim {\mathbb H}^2 \times S^2$,
implies that 
the $SO(3)$-invariant $SU(2)$ instantons are reduced by 
an $S^2$-dimensional reduction to 
$U(1)$ Abelian-Higgs vortices on a hyperbolic plane ${\mathbb H}^2$ 
with a specific curvature \cite{Witten:1976ck}.
Here, we extend this relation to the non-Abelian case \cite{Manton:2010wu}. 

First, let us consider the generators of the $SU(2N)$ gauge group, 
which are invariant under the diagonal group of 
the spatial rotation $SO(3)$ and $SU(2) \subset SU(2N)$ 
generated by $\mathbf 1_N \otimes \sigma_i$. 
It is convenient to take the following basis for the $SU(2)$-invariant generators 
\beq
\Lambda = T \otimes P, \hs{10} 
\tilde \Lambda = \tilde T \otimes \tilde P, 
\eeq
where $T$ and $\tilde T$ are $N$-by-$N$ Hermitian matrices that can be viewed as the generators of $S[U(N) \times U(N)]$. The 2-by-2 matrices $P$ and $\tilde P$ are the projection operators defined by
\beq
P \equiv \frac{\mathbf 1_2 - \hat x_i \sigma_i}{2}, \qquad 
\tilde P \equiv \frac{\mathbf 1_2 + \hat x_i \sigma_i}{2},
\eeq
where $\sigma_i~(i=1,2,3)$ denote the Pauli matrices.
We define the complex coordinate on $\mathbb H^2$ by $z \equiv x^4 + ir$ with 
$r \equiv  \sqrt {(x^1)^2+ (x^2)^2+(x^3)^2}$, whereas 
a unit vector for $S^2$ is denoted by $\hat x_i \equiv  x_i/r$ .

The general $SU(2)$-invariant gauge one-form on $\R^4$ takes the form
\beq
A_{4d} &=& 
A \otimes P + \tilde A\otimes \tilde P 
- \frac{1}{2} (H - \mathbf 1_N) \otimes \omega 
- \frac{1}{2} (H^\dagger - \mathbf 1_N) \otimes \omega^\dagger ,
\label{eq:instanton->vortex}
\eeq
where $\omega$ is the $SU(2)$-invariant one-form on $S^2$ defined by
\begin{eqnarray}
\omega ~\equiv~ i P \sigma_i d \hat x_i ~=~ i \sigma_i d \hat x_i \tilde P ~=~
\frac{i}{2r} \left( \delta_{ij} - \hat x_i \hat x_j - i\epsilon_{ijk} \hat x_k \right) \sigma_j dx_i.
\end{eqnarray} 
As we will see, the one-forms $A$ and $\tilde A$ can be interpreted as 
the gauge fields on $\mathbb H^2$. 
For the gauge field \eqref{eq:instanton->vortex}, 
the field strength $F_{4d} = dA_{4d} +i A_{4d} \wedge A_{4d}$ is given by
\begin{eqnarray}
F_{4d} &=& 
F \otimes P + \tilde F \otimes \tilde P + \frac{1}{2} \D H \otimes \omega
+ \frac{1}{2} \D H^\dagger \otimes \omega^\dagger \nonumber\\
&&\quad -\frac{i}4 (H^\dagger H-{\bf 1}_N)\otimes \omega^\dagger \wedge \omega
-\frac{i}4 (H H^\dagger-{\bf 1}_N)\otimes \omega \wedge \omega^\dagger. 
\end{eqnarray}
Substituting this field strength into the Yang-Mills action and integrating over $S^2$, we find that the 4d Yang-Mills action reduces to the action 
of the 2d $U(N) \times U(N)$ gauge theory for the integrable case (\ref{eq:integrable})
\begin{eqnarray}
\frac{1}{g_4^2} \int {\rm Tr} [ F_{4d} \wedge \ast F_{4d}] 
= \{ \mbox{Eq.\,(\ref{eq:action}) with Eq.\,\eqref{eq:integrable}} \} , \hs{10}
v^2=\frac{4\pi}{g_4^2},
\end{eqnarray}
where we have used the following relations 
\beq
\ast (dz\wedge d\bar z) \otimes \mathbf 1_2 &=& r^2 ( \omega^\dagger \wedge \omega - \omega \wedge \omega^\dagger), \\
\ast (dz \wedge \omega) &=& -dz \wedge \omega, \\
\ast(d\bar z\wedge \omega) &=& \phantom{-} d\bar z\wedge \omega, \\
\int_{S^2}\frac{i}{2} {\rm tr}[\omega\wedge \omega^\dagger] &=& 4\pi. 
\eeq
Similarly, the topological charge of instantons in 4d reduces to
that of vortices in 2d 
\beq
-\frac{1}{g^2_{4d}} \int_{\mathbb R^4} \tr \left[ F \wedge F \right] = - v^2 \int_{\mathbb H^2} (F - \tilde F).
\eeq
This implies that the anti-self-dual equation, $F_{4d} = - \ast F_{4d}$, 
for Yang-Mills instantons reduces to the BPS vortex equations 
(\ref{eq:BPSeq1}), (\ref{eq:BPSeq2}), and (\ref{eq:BPSeq3})
on a hyperbolic plane ${\mathbb H}^2$ 
for the integrable case (\ref{eq:integrable}). 
Therefore, we can use $SU(2N)$ instanton solutions to 
obtain $S[U(N)\times U(N)]$ vortex solutions.

\subsection{$SU(2N)$ instantons from the ADHM construction}
In order to construct $SO(3)$-invariant instanton solutions, we use the 
ADHM construction \cite{Atiyah:1978ri}.
Let $B_1$ and $B_2$ be  $k \times k$ complex matrices 
and $I$ and $J$ be $k \times 2N$ and $2N \times k$ complex matrices, 
respectively. 
Then, ``the zero-dimensional Dirac operator" is defined by 
\beq
\nabla^\dagger ~=~ 
\ba{c|cc} 
I & z_2 - B_2 & z_1 - B_1 \\ 
J^\dagger & - (\bar z_1 - B_1^\dagger) & \bar z_2 - B_2^\dagger \ea,
\eeq
where we have defined $z_1 \equiv i x_1 + x_2,~z_2 \equiv x_4 + i x_3$. 
Now, let us consider the following $SO(3)$ action on the ADHM data $(B_i,I,J)$
\beq
\nabla^\dagger ~\rightarrow~ g \nabla^\dagger h, \hs{15} g = \mathbf 1_k \otimes U^\dagger, \hs{5} h = {\arraycolsep=1mm \ba{c|c} \mathbf 1_N \otimes U & \\ \hline & \mathbf 1_k \otimes U \ea},
\eeq
where $U$ is an arbitrary $SU(2)$ matrix. The most general ADHM data $(B_i,I,J)$ which are invariant under the $SO(3)$ transformation take the following forms (see Appendix \ref{appendix:inv_data})
\beq
B_1=B_1^\dagger=0, \hs{10} B_2=B_2^\dagger=T, \hs{10} \ba{c} I \\ J^\dagger \ea = \ba{c|c} \psi & 0 \\ \hline 0 & \psi \ea
= \psi \otimes \mathbf 1_2, \label{eq:assumption}
\eeq
where $T$ is an arbitrary $k$-by-$k$ Hermitian matrix and $\psi$ is an arbitrary $k$-by-$N$ matrix. We can show that the $SO(3)$-invariant ADHM data automatically satisfy the following ADHM equations. 
\beq
0 = [B_1, B_1^\dagger] + [B_2,B_2^\dagger] + II^\dagger - J^\dagger J, 
\quad\quad
0 = [B_1, B_2] + I J.
\eeq
More generally, the ADHM equations are satisfied if $B_1$ and $B_2$ are diagonal and $(I,J^\dagger)$ take the form \eqref{eq:assumption}.
In such a case, the operator $\nabla^\dagger$ is given by 
\beq
\nabla^\dagger 
~=~ \ba{cc} \psi \otimes \mathbf 1_2  & (x^\mu \mathbf - T^\mu) \otimes \bar e_\mu \ea, \label{eq:nabla}
\eeq
where $e_\mu = (-i \sigma_i, \mathbf 1_2)$ and $\bar e_\mu = (i \sigma_i, \mathbf 1_2)$, and we have taken $B_1 = i T_1 + T_2$ and $B_2 = i T_3 + T_4$ with $k \times k$ mutually commuting Hermitian matrices $T_{\mu}$. For notational simplicity, first, we deal with the case of the mutually commuting matrices $T_\mu$ and then return to the $SO(3)$-invariant case by setting $T_4 = T$ and $T_i = 0~(i=1,2,3)$.

For the operator $\nabla^\dagger$ of the form \eqref{eq:nabla}, the zero modes $V$, which are a $(2N+2k) \times 2N$ complex matrix satisfying the equation 
\beq 
\nabla^\dagger V=0,  \label{eq:Dirac-instanton}
\eeq
are found to be 
\beq
V = \ba{c} \mathbf 1_N \otimes \mathbf 1_2 \\ 
-(x^\nu - T^\nu) 
\left[ (x^\mu - T^\mu)^2 \right]^{-1} \psi \otimes e_\nu
\ea 
(S^{\dagger-1} \otimes \mathbf 1_2). \label{eq:V}
\eeq
Here, $S$ is an $N$-by-$N$ matrix determined 
from the orthogonality condition
\beq
V^\dagger V &=& \mathbf 1_N, \label{eq:V-orthogoality}
\eeq
or equivalently
\beq
SS^\dagger &=& \mathbf 1_N + \psi^\dagger \left[ (x^\mu - T^\mu)^2 \right]^{-1} \psi, \label{eq:SSdagger-instanton}
\eeq
where we have used the identity $\bar e_\mu e_\nu+\bar e_\nu e_\mu= 2\delta_{\mu\nu} \mathbf 1_2$. From the matrix $V$, 
the instanton solutions can be explicitly given by
\beq
A_{4d} ~=~ - i V^\dagger d V ~=~ - \frac{i}{2} S^{-1} \p^\mu S \otimes \left( \delta_{\mu \nu}  \mathbf 1_2- i \eta^{(+)}_{\mu \nu} \right) dx^\nu + ({\rm h.c.}), 
\eeq
where $\eta_{\mu \nu}^{(+)}$ is the self-dual 't Hooft tensor defined by
$\eta_{\mu \nu}^{(+)}
=\frac{1}{2i}\left(\bar e_\mu e_\nu-\bar e_\nu e_\mu\right)$.
This solution can be viewed as a generalization of the 't Hooft's multi-instanton configuration for the $SU(2)$ gauge group. 

For our purpose, we impose the $SO(3)$ invariance by setting
$T_4 = T$ and $T_i = 0~~(i=1,2,3)$.
In this case, Eq.~(\ref{eq:SSdagger-instanton}) indicates that matrix $S$
is independent of the coordinates of $S^2$.
Thus, the solutions become 
\begin{eqnarray}
A_{4d} =
  i \left[- S^{-1} \p_{\bar z} S \otimes P + 
\p_{\bar z} S^\dagger S^{\dagger-1} \otimes \tilde P \right] d \bar z 
-i r \left( S^{-1} \p_z S + \p_z S^\dagger S^{\dagger-1} \right)
 \otimes \omega  + ({\rm h.c.}),\quad 
\label{eq:SU(2)inv-instanton}
\end{eqnarray}
where we have used
\beq
(\delta_{\mu\nu}\mathbf 1_2-i\eta_{\mu\nu}^{(+)}) dx^\nu \p_\mu 
&=& (\bar e_\mu d x^\mu) e^\nu \p_\nu \notag \\
&=& 2(dz P +r\, \omega)\p_z+2(d\bar z \tilde P
  -r\, \omega^\dagger )\p_{\bar z}+{\rm derivatives ~on~} S^2.
\eeq


\section{All Exact $S[U(N) \times U(N)]$ Vortex Solutions}
\label{sec:hyperbolic-solution}
\subsection{Exact solutions}
Comparing Eq.~(\ref{eq:instanton->vortex}) with Eq.~(\ref{eq:SU(2)inv-instanton}), 
we can obtain the vortex solutions $A_z$ and $H$. 
Let $T$ be a $k \times k$ Hermitian matrix and 
$\psi$ be a $k \times N$ complex matrix,  
made of the respective moduli parameters.
The general form of the vortex solution is
\beq
A_\alpha = - i W^{\dagger} \p_\alpha W, \hs{10}
\tilde A_\alpha = - i \tilde W^{\dagger} \p_\alpha \tilde W, \hs{10}
H = W^{\dagger} \tilde W,\hs{10} (\alpha = z,\bar z), \label{eq:solution}
\eeq
where $W$ and $\tilde W$ are $(N+k)\times k$ matrices, given by
\beq 
W \equiv \ba{c} \mathbf 1_N \\ (\bar z - T)^{-1} \psi \ea S^{\dagger-1}, \hs{10} 
\tilde W \equiv \ba{c} \mathbf 1_N \\ (z - T)^{-1} \psi \ea S^{\dagger-1} \label{eq:V-tildeV}
\eeq
with $S(z,\bar z)$ satisfying 
\beq
SS^\dagger ~=~ \mathbf 1_N + \psi^\dagger 
\left[ (z - T)(\bar z -T) \right]^{-1} \psi. \label{eq:SSdagger-vortex}
\eeq
Here, $S$ is the same matrix as the one for instantons in Eq.~(\ref{eq:V}); 
condition (\ref{eq:SSdagger-vortex}) originates from 
Eq.~(\ref{eq:SSdagger-instanton}) with the identification 
$z=x^4 + ir$ and $r^2 = (x^1)^2 + (x^2)^2 + (x^3)^2$.

\subsection{The ADHM construction for vortices}
From the fact that the $(N+k)\times k$ matrices $W$ and $\tilde W$ 
in Eq.~(\ref{eq:V-tildeV}) are analogous to 
the $2(N+k) \times 2k$ matrix $V$ in Eq.~(\ref{eq:V}) 
for the ADHM construction of instantons, 
the solutions can be recast into the ADHM form.
In fact, for a given ADHM date $(T,\psi)$, the matrices $W$ and $\tilde W$ 
are solution of the ``Dirac equations''
\beq
\nabla_{v}^{\dagger} W = 0, \hs{10} 
\tilde \nabla_{v}^{\dagger} \tilde W = 0,
\eeq
where the ``Dirac operators'' for the vortices are given by
\beq
\nabla_{v}^{\dagger} \equiv \ba{cc} \psi & T - \bar z \ea, \hs{10} 
\tilde \nabla_{v}^{\dagger} \equiv \ba{cc} \psi & T -z\ea.
\label{eq:Dirac-op-vor}
\eeq
These Dirac operators are analogous to those of 
instantons in Eq.~(\ref{eq:Dirac-instanton}). 
Condition (\ref{eq:SSdagger-vortex}) is equivalent to 
the orthogonality conditions for matrices $W$ and $\tilde W$:  
\beq
W^{\dagger} W = \mathbf 1_N, \hs{10} 
\tilde W^{\dagger} \tilde W = \mathbf 1_N.
\eeq
These conditions are also counterparts of 
Eq.~(\ref{eq:V-orthogoality}) for instantons.

We need to check the existence of the inverse of $\Delta^\dagger \Delta$ in the ADHM construction 
for instantons, 
and therefore there should be a corresponding condition for the operators in Eq.~(\ref{eq:Dirac-op-vor}) for vortices. However we will not study it in more detail in this paper 
and leave it as a future problem.


\subsection{Moduli space}
Here, we discuss the moduli parameters encoded in 
solutions (\ref{eq:solution}), (\ref{eq:V-tildeV}), and (\ref{eq:SSdagger-vortex}) and identify the moduli space.  
The solutions have the following redundancy in the moduli data $(T,\psi)$:
\beq
 T \to U T U^{-1} , \quad \psi \to U \psi, \quad U \in U(k) .
\label{eq:U(k)}
\eeq
They can be fixed as
\beq
T = \ba{cccc} t_1 & & & \\ & t_2 & & \\ & & \ddots & \\ & & & t_k \ea, \hs{10}
\psi = \ba{cccc} \rho_{11} & \rho_{12} & \cdots & \rho_{1N} \\ \rho_{21} & \rho_{22} & \cdots & \rho_{2N} \\ \vdots & & \ddots & \vdots \\ \rho_{k1} & \rho_{k2} & \cdots & \rho_{kN}\ea, 
\quad t_i \in {\mathbb R}, \quad 
\rho_{ij} \in {\mathbb C} .
\eeq
The remaining $U(1)^k$ redundancy,  
$\rho_{ij} \to \exp (i \theta_i) \rho_{ij}$, 
can be fixed as $\rho_{ii} \in {\mathbb R}$. 
Therefore, the dimension of the $S[U(N) \times U(N)]$ vortex moduli space is 
1/4 of that of the (framed) $SU(2N)$ instanton moduli space:
\beq
\dim_{\mathbb R} {\cal M}_{k,N}^{\rm vortex} = 2k N.
\eeq
If we further divide the moduli space 
by the $SU(N)$ (global) gauge symmetry $\psi \rightarrow \psi g,~g \in SU(N)$,
the dimension of the moduli space becomes
\beq
 \dim_{\mathbb R} [{\cal M}_{k,N}^{\rm vortex}/SU(N)] = 
\left\{
\begin{array}{cc}
 2 k N-N^2+1&  {\rm for~}k> N\\
 k^2+1 & {\rm for~} k\le N
\end{array}\right..
\eeq

To determine the physical meaning of the moduli parameters, 
let us calculate
\beq
 \det H &\propto& \det \left({\mathbf 1}_N + \psi^\dagger (z{\mathbf 1}_k  - T)^{-2} \psi\right) \nonumber \\
 &\propto& \det \left({\mathbf 1}_k 
   + (z{\mathbf 1}_k  - T)^{-2} \psi \psi^\dagger\right)
\nonumber\\
 &\propto& \det \left((z{\mathbf 1}_k  - T)^2 + \psi \psi^\dagger\right). 
\label{eq:det_H}
\eeq
Now, let us define a $k\times k$ complex matrix $Z$ by 
\beq
 Z \equiv T + iR 
\eeq
with a $k \times k$ Hermitian matrix $R$ satisfying
\beq
  i [T,R] + \psi\psi^\dagger = R^2. \label{eq:R}
\eeq
Hence, Eq.\,\eqref{eq:det_H} can be rewritten as
\beq
 \det H &\propto& \det (z - Z) (z- Z^\dagger) . \label{eq:det-H}
\eeq
Since the unbroken gauge symmetry becomes larger inside the vortex cores, 
the zeros of $\det H$ can be interpreted as the vortex positions. 
Therefore, Eq.~(\ref{eq:det-H}) implies that 
the eigenvalues of $Z$ are the vortex positions.
It is interesting to see that Eq.~(\ref{eq:R}) 
is a remnant of the D-term condition of 
a K\"ahler quotient construction of the vortex moduli space.
We thus obtain 
\beq
  {\cal M}_{k,N} \simeq 
\left\{ (Z,\psi)\ \bigg| \ {1\over 2} [Z^\dagger,Z] + \psi\psi^\dagger = R^2 
 \right\}/U(k), \label{eq:Kahler-quotient}
\eeq
where the $U(k)$ action is 
$Z \to U Z U^{-1}$, $\psi \to U\psi$, and $R \to U R U^{-1}$, 
as in Eq.~(\ref{eq:U(k)}). 
This is analogous to the K\"ahler quotient 
for $U(N)$ vortices on flat space ${\mathbb C}$ \cite{Hanany:2003hp}.
The moduli $Z$ represent the vortex positions, 
and the moduli $\psi$ can be identified as the orientational moduli, 
which are $k$ copies of ${\mathbb C}P^{N-1}$    
for the separated vortices. 
The difference between our hyperbolic case and the flat case 
is that the right hand side of the D-term condition is $R^2$ 
in our case while it is just $(4\pi/g^2){\mathbf 1}_k$ for the flat case. 
Since the eigenvalues of $R$ are
the vortex positions in the $r$ coordinate, 
our D-term condition can be interpreted as 
a result of a position-dependent gauge coupling, 
as can be inferred from Eqs.\,\eqref{eq:action} and (\ref{eq:sigma}). 
Note that the K\"ahler quotient in 
Eq.~(\ref{eq:Kahler-quotient}) does not give 
the correct moduli space metric, 
as in the case of the flat space.

Although the $\C P^{N-1}$ orientational moduli for a single vortex can be absorbed by a global gauge transformation,  the relative orientations change the physical quantities. 
Let us consider a coincident vortex configuration in the $N=k=2$ case. 
If we set the matrices $T$ and $\psi$ as 
\begin{eqnarray}
T = \left(
\begin{array}{cc}
a & 0 \\ 
0 & -a
\end{array}\right),\quad 
\psi = \left(
\begin{array}{cc}
\sqrt{r^2_0+a^2} & 0 \\
2a & \sqrt{r^2_0-3a^2}
\end{array} \right), \hs{10} 
a \in \left[0, \frac{r_0}{\sqrt{3}} \right], 
\label{eq:coincident}
\end{eqnarray}
the matrix $R$ is solved as
\begin{eqnarray}
R = \left(
\begin{array}{cc}
r_0 & \frac{r_0+i a}{\sqrt{r_0^2+a^2}} a \\
\frac{r_0-ia}{\sqrt{r_0^2+a^2}} a & r_0 
\end{array} \right).
\end{eqnarray}
In this setting, the matrix $Z = T + i R$ has the degenerate eigenvalue $i r_0$; 
hence, the two vortices are coincident. 
Since the vortex position is independent of $a$, 
$a$ parameterizes the internal orientation of the vortices. 
Indeed, we can see from Eq.~(\ref{eq:coincident}) that 
$(T,\psi)$ reduce to two copies of the data for an Abelian vortex at $a=0$ 
while $(T,\psi)$ become identical to the data of two vortices in the Abelian case 
at $a=r_0/\sqrt{3}$. 
We can confirm that the parameter $a$ is physical by observing the trace of the magnetic flux $F_{z \bar z} = - \tilde F_{z \bar z}$.
\beq
i \sigma^{-1} \tr \, F_{z \bar z} = 8 r^2 (a^2 + r_0^2) \left[ \frac{1}{\big\{ |z|^2 + r_0^2 + a (z- \bar z) \big\}^2} + \frac{1}{\big\{ |z|^2 + r_0^2 - a (z- \bar z) \big\}^2} \right]. 
\eeq

\section{Summary and Discussion}  \label{sec:conclusion}
In summary, we have constructed a complete set of 
all the exact solutions of non-Abelian vortices 
in the $S[U(N)\times U(N)]$ gauge theory with a bi-fundamental scalar field 
on a hyperbolic plane with a certain curvature, 
by using $SO(3)$-invariant $SU(2N)$ Yang-Mills instanton solutions. 
We also have established the ADHM construction for non-Abelian vortices. 
We further identified the complete moduli space of $k$ vortices, 
consisting of the moduli parameters encoded in 
the $k \times k$ matrix $Z$ for the position moduli and 
the $k\times N$ matrix $\psi$ for the orientational moduli.
We have found the K\"ahler quotient for the moduli space, 
whose complex dimension is $kN$ as in the flat case. 

Future works on related topics will include studies on 
the index theorem of vortices in quiver gauge theories, 
the explicit moduli space metric, 
and low-energy dynamics of vortices;
and an extension to arbitrary gauge groups \cite{Eto:2008yi},
particularly $SO(N)$ and $USp(2N)$ \cite{Eto:2009bg}, 
from the Yang-Mills instantons with corresponding groups. 
Since the hyperbolic surface is topologically equivalent to the flat space, quantum dynamics such as the vortex counting should be studied on the hyperbolic surface.
In the case of $N=1$, our model reduces to the Abelian-Higgs model \cite{Witten:1976ck}, in which the vortex equation is reduced to the Liouville equation. 
This implies the presence of a non-Abelian generalization of  the Liouville equation. 

\section*{Acknowledgments}

This work is supported in part by 
Grant-in-Aid for Scientific 
Research, Nos.~23740226 (M.E.) and 20740141 (M.N.),  
from the Ministry of Education, 
Culture, Sports, Science and Technology (MEXT) of Japan,
by the ``Topological Quantum Phenomena'' 
Grant-in Aid for Scientific Research 
on Innovative Areas, No. 23103515 (M.N.),   
and by Japan Society for the Promotion of Science (JSPS),
Academy of Sciences of the Czech Republic (ASCR) under
the Japan - Czech Republic Research Cooperative Program
(M.E.).

\appendix
\section{Hyperbolic plane}\label{sec:hyperbolic}
The hyperbolic plane is a subspace in $\R^{2,1}$ given by
\beq
X_1^2 + X_2^2 - X_3^2 = - R^2.
\eeq
The solution is parameterized by $\phi \in [0,2\pi), \rho\in \mathbb
R_{\ge 0}$ as 
\beq
X_1 = R \cos \phi \sinh \rho, \hs{10} X_2 = R \sin \phi \sinh \rho, \hs{10} X_3 = R \cosh \rho.
\eeq
The metric is given by
\beq
ds^2 ~=~ d X_1^2 + d X_2^2 - d X_3^2 ~=~ R^2 (d \rho^2 + \sinh^2 \rho \, d \phi^2).
\eeq
which gives a constant scalar curvature $-1/R^2$.
The hyperbolic plane can be parametrized by a complex coordinate $y$ in
an unit disc as
\beq
y = \tanh \frac{\rho}{2} e^{i \phi}, \hs{10} |y|<1.
\eeq
Then, the metric becomes
\beq
d s^2 ~=~ 2 g_{y \bar y} dy d \bar y ~=~ \frac{4 R^2}{(1-|y^2|)^2} d y d \bar y.
\eeq
An upper-half plane is also used to parameterize the hyperbolic plane,
where a complex coordinate $z$ is given by
\beq
z = \frac{y+i}{1 + i y}, \hs{10} {\rm Im} \, z > 0 .
\eeq
In terms of this coordinate the metric becomes
\beq
ds^2 = R^2 \frac{dz d\bar z}{({\rm Im} \, z)^2} .
\eeq
The $SO(2,1)$ isometry of $\mathbb R^{2,1}$ acts on $z$ as
\beq
z \rightarrow \frac{a z + b}{c z + d}, \hs{10} a,b,c,d \in \R.
\eeq

\section{$SO(3)$-invariant ADHM data}\label{appendix:inv_data}
In this section, we show that the most general $SO(3)$-invariant ADHM data takes the form of Eq.\,\eqref{eq:assumption}. First, let us rewrite the pair of the matrices $(I,J^\dagger)$ as
\beq
\ba{c} I \\ J^\dagger \ea = \ba{c|c} \phi_4 + i \phi_3 & i \phi_1 + \phi_2 \\ \hline i \phi_1 - \phi_2 & \phi_4 - i \phi_3 \ea = \phi_i \otimes \sigma_i + \phi_4 \otimes \mathbf 1_2,
\eeq
where $\phi_i~(i=1,2,3)$ and $\phi_4$ are arbitrary $k$-by-$N$ matrices.
The $SO(3)$ transformations rotate $\phi_i$ and $\phi_4$ as
\beq
\phi_i \rightarrow R_i{}^j \phi_j, \hs{10} \phi_4 \rightarrow \phi_4, \hs{10} R^T R = \mathbf 1_3.
\eeq
This transformation can be canceled if there exists a $U(k)$ gauge transformation such that
\beq
U \phi_i = R^j{}_i \phi_j, \hs{10} U \phi_4 = \phi_4, \hs{10} U \in U(k).
\eeq
This implies that the matrix $U$ must be in a direct sum of several triplet and singlet representations: $\phi_4 = 0$ for the triplet part while $\phi_i = 0$ for the singlet part. However, we can show that the triplet part does not satisfy the ADHM equations
\beq
\frac{1}{2} \epsilon_{ijk} [T_j,T_k] + [T_i,T_4] + \frac{1}{2} \epsilon_{ijk} (\phi_j \phi_k^\dagger - \phi_k \phi_j^\dagger) + \phi_i \phi_4^\dagger - \phi_4 \phi_i^\dagger = 0.
\label{eq:adhM}
\eeq
For simplicity, let us consider the case of a single triplet representation ($k=3$). The explicit form of the invariant data is given by
\beq
(T_i)_{rs} = i a \epsilon_{irs}, \hs{10} (T_4)_{rs} = b \delta_{rs}, \hs{10} (\phi_i)_{rA} = \delta_{ri} \psi_A, \hs{10} \phi_4 = 0,
\eeq
where $a$, $b \in \R$ and $\psi_A \in \C ~(A=1,\cdots,N)$ are arbitrary parameters. Then, Eq.\,\eqref{eq:adhM} reduces to
\beq
\epsilon_{irs} (a^2 + |\psi_A|^2 ) = 0.
\eeq 
This allows only a trivial solution $a=\psi_A=0$ for which 
the operator $\Delta^\dagger\Delta$ is not invertible at
$x^\mu=(0,0,0,b)$. 
Therefore, there is no solution to the ADHM equations for the triplet representation. Similarly, we can in general show that there is no $SO(3)$-invariant ADHM data in a direct sum representation containg the triplet representation. Namely, the most general form of the $SO(3)$-invariant data should be in a direct sum of the singlet representations:
\beq
T_i = \phi_i = 0~~(i=1,2,3), \hs{10} T_4 = T, \hs{10} \phi_4 = \psi, 
\eeq
where $T$ and $\psi$ are arbitrary $k$-by-$k$ and $k$-by-$N$ matrices, respectively.


\end{document}